\documentclass[12pt]{iopart}
\usepackage[dvips]{epsfig}

\begin{document}
\title{Short and long range correlated motion observed
in colloidal glasses and liquids}
\author{Eric R.~Weeks$^{\rm (1)}$, John 
C.~Crocker$^{\rm (2)}$ and D.~A.~Weitz$^{\rm (3)}$}
\address{$^{\rm (1)}$Physics Department, Emory University,
Atlanta, GA 30322}
\ead{weeks@physics.emory.edu}
\address{$^{\rm (2)}$Department of Chemical and Biomolecular
Engineering, 
University of Pennsylvania, Philadelphia, PA 19102}
\address{$^{\rm (3)}$Department of Physics and DEAS, Harvard
University, Cambridge, MA 02138}
\date{\today}

\begin{abstract}
We use a confocal microscope to examine
the motion of individual particles in a dense colloidal
suspension.  Close to the glass transition, particle
motion
is strongly spatially correlated.  The correlations decay exponentially
with particle separation, yielding a dynamic length scale
of $O(2-3\sigma$) (in terms of particle diameter $\sigma$).
This length scale grows modestly as the glass transition is
approached.
Further, the correlated motion exhibits a strong
spatial dependence on the pair correlation function $g(r)$.
Motion within glassy samples is weakly correlated, but with a
larger spatial scale for this correlation.
\end{abstract}



\maketitle

\section{Introduction}

The viscosity of a glass-forming material increases rapidly by
many orders
of magnitude as it is cooled, without any corresponding
structural change to account for the viscosity change
\cite{reviews1,reviews2}.  Adam and
Gibbs suggested that a growing dynamic length
scale may relate to the viscosity growth \cite{gibbs}.  Their idea
has been interpreted in several ways
\cite{reviews1,reviews2,kivelson94,gotze92,ngai98,parisi99,garrahan03},
but no experiment has been able to directly observe
dynamic length scales, nor is it clear what form these length scales
would take.  Indirect evidence is provided by
experiments which locally perturb materials in a variety of
ways, and observe the relaxation response of the material
and its dependence on the scale of the perturbation
\cite{ediger00}.
Further evidence for a dynamic length scale comes from
experiments done in thin films or small pores \cite{mckenna05}, but
these experiments are unable to directly observe the nature
of any correlated motion.  Recent simulations have examined
correlation functions in systems of Lennard-Jones particles
\cite{donati99}
and hard spheres \cite{doliwa00} finding evidence for spatial correlations
and a possible dynamical correlation length scale
\cite{doliwa00,donati99}.

We study a system of colloidal particles which interact only via
repulsive forces, and which have a glass transition as their
concentration is increased.  We use a confocal microscope to
track the motions of several thousand particles for several
hours, long enough for most particles to undergo nontrivial
displacements.  The mobilities of these particles
(the magnitudes of their displacements) are correlated over
distances of $\sim 2-3\sigma$, in terms of the particle diameter
$\sigma$.  These correlations reflect large-scale
cooperative rearrangements of particles seen previously
\cite{weeks00,kegel00,doliwa00,donati99,poole98}.  We also examine
the correlations between the directions of particle displacements.
While the directions are correlated, this does
not appear to be as strong a signature of the rearrangements.
These measurements are direct experimental studies
of the nature of the long-range correlations present near a
glass transition, and indicate that rearranging regions of
particles are composed of particles with large displacements,
but which do not all move in similar directions; rather,
these regions are internally rearranging \cite{weeks01}.

\section{Experimental Methods}

Our system consists of a suspension of colloidal
poly-methyl-methacrylate, sterically stabilized and dyed with
a rhodamine dye \cite{dinsmore01}.  These particles
are slightly charged, and have a hard-sphere 
diameter $\sigma=2.36$ $\mu$m and a polydispersity of 5\%.
They are suspended in a mixture
of cycloheptylbromide and decalin which nearly index and
density matches the particles.  The samples are prepared with
a constant volume fraction $\phi$ and sealed into microscope
chambers.  We observe crystallization occurring at $\phi =
0.42$, slightly lower than the value expected for hard spheres
($\phi_{HS}=0.494$) \cite{pusey86a,pusey86b}.  Samples with
$\phi>\phi_g \approx 0.58$ do not form crystals within the
bulk even after sitting at rest for several months, allowing
us to identify $\phi_g$ as the glass transition volume
fraction, in agreement with previous work \cite{pusey86a,pusey86b}.
Prior to observation, samples with crystals
in them are shear-melted by means of a stir bar; there is
a reasonable separation of time scales between the decay of
transient flows caused by the stirring ($< 20$ min) and the
onset of crystallization for these samples ($> 5$ - 10 hours,
as determined by methods described in \cite{gasser01}).

We use confocal microscopy to observe the particle motion.
A single
three-dimensional image is acquired in 10~s, and over the course
of an experiment we acquire several hundred images.
As the mean square displacement curves show
in figure \ref{fig-dt}(a), particles do not move significant
distances on a time scale of 10 s.  We post-process the data
to determine particle positions with an accuracy of 0.03
$\mu$m horizontally and 0.05 $\mu$m vertically \cite{dinsmore01}.
Each three-dimensional image is 69 $\times$ 65 $\times$ 14
$\mu$m and contains several thousand particles.  We focus at
least $25$~$\mu$m from the coverslip to avoid interference from
the wall, and in fact the observed motion appears isotropic.

\section{Results}

\smallskip
\begin{figure}
\centerline{
\epsfxsize=8truecm
\epsffile{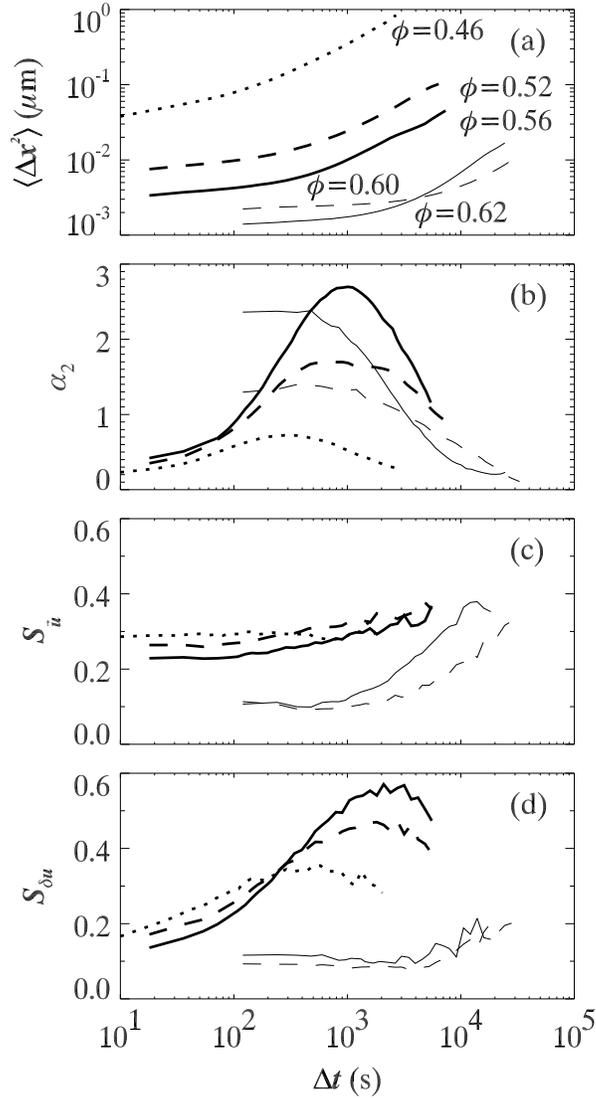}}
\smallskip
\caption{(a) Mean square displacement for colloidal
``supercooled'' fluids (thick lines) and colloidal glasses,
with volume fractions $\phi$ as labeled.  (b) Nongaussian
parameter $\alpha_2$ for the samples shown in (a).  (c,d) 
Vector correlation function $S_{\vec u}(\Delta t)$ and
scalar correlation function $S_{\delta u}(\Delta t)$ for the
samples shown in (a), using $\Delta r$ corresponding to the
first peak of the peak of $g(r)$ for each sample.
}
\label{fig-dt}
\end{figure}

To characterize the behaviour of the samples, we calculate
the particles' mean square displacement $\langle \Delta x^2
\rangle$, shown in figure \ref{fig-dt}(a) for five samples.
These curves all show a plateau,
due to cage-trapping: each particle is confined in a cage
formed by its neighbors.  At longer times, $\langle \Delta
x^2 \rangle$ shows an upturn, indicating that at least
some particles have moved.  For the supercooled fluids
(thick lines, $\phi<\phi_g$), previous work
showed that these motions
correspond to a small subset of particles undergoing cage
rearrangements \cite{weeks00,kegel00,donati99,poole98}.  These particles move
significantly farther than the majority of the particles,
and thus distributions of the particle displacements show
broad tails at these time scales \cite{weeks00,kegel00,donati99,poole98}.
This is quantifiable by the nongaussian parameter
\begin{equation}
\alpha_2(\Delta t) = {\langle \Delta x^4 \rangle \over
3 \langle \Delta x^2 \rangle^2} - 1
\end{equation}
where the moments of $\Delta x$ are calculated from the measured
distributions of the one-dimensional displacements $\Delta x$.
$\alpha_2$ is zero for a gaussian, and larger when the
distributions are broader than a gaussian.  We plot $\alpha_2$
in figure \ref{fig-dt}(b), finding that for supercooled fluids
(thick lines) $\alpha_2$ has a peak corresponding to the end
of the plateau of $\langle \Delta x^2 \rangle$, due to the
presence of the anomalously mobile particles \cite{weeks00}.
For colloidal glasses [thin lines in figure \ref{fig-dt}(a)],
the upturn in $\langle \Delta x^2 \rangle$ is due to aging,
and occurs at a time scale $\Delta t$ which varies with the
time since sample preparation \cite{courtland}.  It is unclear if the motions
responsible for the upturn are due to cage rearrangements
\cite{weeks00}.

Intriguingly, in supercooled colloidal fluids, the motion
of these cage-rearranging particles is spatially localized,
and the peak of $\alpha_2(\Delta t)$ corresponds to the
existence of large clusters of these particles all moving
simultaneously \cite{weeks00}.  In fact, the positions of
mobile particles appear in localized clusters over a range
of time scales $\Delta t$, and the $\Delta t$ dependence of
the typical cluster size appears qualitatively similar to the
$\Delta t$ dependence of $\alpha_2$ \cite{weeks00}.

We wish to determine how these previously observed clusters of
cage-rearranging particles are manifested in correlations of
the motion of individual particles.  In a dilute suspension
of particles, particles have only hydrodynamic interactions,
and the correlations between the displacement vectors
$\vec u$ of two particles should decay as 
$1/\Delta r$ with increasing separation $\Delta r$
\cite{batchelor76,crocker97,crocker00}.  This result would still hold
true for dilute tracers in any homogeneous viscoelastic medium
\cite{levine00a,levine00b}.  However, it seems unlikely that our samples
are homogeneous on the length scales we consider; our imaging
window has a width of only $\sim 25$ particle diameters.
The direct interparticle forces (their electric charge, and
steric repulsion) are likely to be as important as hydrodynamic
interactions.

Thus, we do not necessarily expect $1/\Delta r$ decay of
correlations, especially given the apparently cooperative and
localized nature of the cage rearrangement motion.  For example,
a simulation of hard spheres at large volume fractions found
exponential decay \cite{doliwa00}.  This suggests particles may
rearrange by translating together in a group.  One possibility
for this group motion is that cage rearranging particles move
in parallel directions \cite{weeks00,kegel00,doliwa00,donati99}.  An
alternate possibility is that {\it mobility} may be correlated
over long distances.  Mobility is the magnitude of particle
displacements $u = |\vec u|$, sometimes with the average
subtracted off ($\delta u \equiv u - \langle u \rangle$).
In fact, the previous observations of clusters of anomalously
mobile particles are a direct indication that mobility is
correlated; simulations have provided evidence that there is
a correlation length scale \cite{doliwa00,donati99}.
Correlations of mobility are consistent with both collectively
translating regions, and also internal rearrangements within
localized regions.

To study these possibilities, we compute two correlation
functions from our data, one using $\vec u$ and one using the mobility
$\delta u$ \cite{doliwa00,crocker00,donati99}.  We define
\begin{eqnarray}
S_{\vec{u}}(\Delta r, \Delta t) &=&
{\langle \vec u_i \cdot \vec u_j \rangle \over \langle 
u^2 \rangle},\\
S_{\delta u}(\Delta r, \Delta t) &=&
{\langle \delta u_i \delta u_j \rangle \over \langle
(\delta u)^2 \rangle},
\end{eqnarray}
where the time scale $\Delta t$ is used to define the
displacements $\vec u$ and mobility $\delta u$.
For both formulas the numerator average is over all pairs of
particles $i,j$ with separations $\Delta r$, and the denominator
average is over all particles \cite{cv,cs}.  The normalization
of both of these functions is chosen so that perfectly
correlated motion ($\vec u_i=\vec u_j$ for all particles $i,j$)
corresponds to $S_{\vec u} = S_{\delta u} = 1$, perfectly
anticorrelated motion ($\vec u_i=-\vec u_j$) corresponds to
$S_{\vec u} = -1$, and uncorrelated motion gives $S_{\vec
u}=S_{\delta u}=0$.  Note that $S_{\delta u}$ measures {\it
fluctuations} of mobility, so either anomalously mobile or
anomalously immobile particles would show positive correlation.
By examining the $\Delta r$ dependence of these functions we
learn about spatially correlated motion, and by examining the
$\Delta t$ dependence we study the correlated motion seen at
different time scales.

To study the $\Delta t$ dependence of $S_{\vec u}$ and
$S_{\delta u}$, for convenience we choose a fixed value
$\Delta r = \Delta r_{nn}$ for each data set, set by the
first maximum of the pair correlation function $g(r)$ (which
is slightly larger than at $2\sigma$, and depends slightly on
$\phi$, due to the charging of
the particles).  Pairs of particles with separation $\Delta
r_{nn}$ are nearest neighbors and are strongly correlated
due to their close proximity, as will be shown later.

We plot
$S_{\vec u}(\Delta t,\Delta r_{nn})$ in figure \ref{fig-dt}(c) and
$S_{\delta u}(\Delta t,\Delta r_{nn})$ in figure \ref{fig-dt}(d).
For the supercooled fluids (thick lines), the behaviour of these
two functions is dramatically different.  $S_{\vec u}$ is nearly
constant as a function of $\Delta t$, whereas $S_{\delta u}$
is small at short lag times, rises to a maximum, and then has
a downturn.  Strikingly, the $\Delta t$ dependence of $S_{\delta
u}$ is similar to that of the nongaussian parameter $\alpha_2$,
which can be seen for three supercooled fluid samples
by comparing
the thick lines in figure \ref{fig-dt}(b) and figure \ref{fig-dt}(d).
While $S_{\delta u}$ reaches its maximum at slightly larger
values of $\Delta t$, it appears that the anomalously mobile
particles (which cause the increase of $\alpha_2$) are
directly responsible for the increasing correlation measured
by $S_{\delta u}$.  While
the time dependence of the vectorial correlations
$S_{\vec u}$ is much weaker than for the mobility,
there is a slight rise at larger
$\Delta t$ seen in figure \ref{fig-dt}(c) \cite{doliwa00}.  It has
been conjectured that $S_{\vec u}(\Delta t \rightarrow \infty)$
should be nonzero, as short-lag-time interparticle correlations
will always provide a contribution to $\langle \vec u_i \cdot \vec
u_j \rangle$ even if particles are uncorrelated at longer
lag times \cite{doliwa00}.  This may explain why we do not
see a downturn in $S_{\vec u}$ at large $\Delta t$.

Figures \ref{fig-dt}(c,d) also show that at short lag times,
$S_{\vec u}>S_{\delta u}$, although at the cage-breaking time
scales, the opposite is true.  These results, along with the
similar lag time dependence of $\alpha_2$ and $S_{\delta u}$,
suggest that the cage rearrangements are due to mobility
correlations, rather than a strong directional correlation.
The implied picture is that regions of cage rearrangements
are composed of highly mobile particles, which move in many
directions.  While the motions of neighboring particles are
somewhat directionally correlated, the cage rearrangements
reflect regions of internal restructuring rather than
large-scale cooperative translations.

Further confirmation of this picture comes from the volume
fraction dependence of the correlation functions, seen in
figure \ref{fig-dt}(c,d).  As $\phi$ increases towards
the glass transition at $\phi_g \approx 0.58$, $S_{\vec
u}$ changes only slightly, while $S_{\delta u}$ changes
dramatically, again similar to the behaviour of $\alpha_2$.
The growth of $S_{\delta u}$ (and non-growth of $S_{\vec
u}$) indicate that as the glass transition is approached,
mobility correlations become increasingly important, while
the directional correlations remain virtually unchanged.

\smallskip
\begin{figure}
\centerline{\epsfxsize=8truecm\epsffile{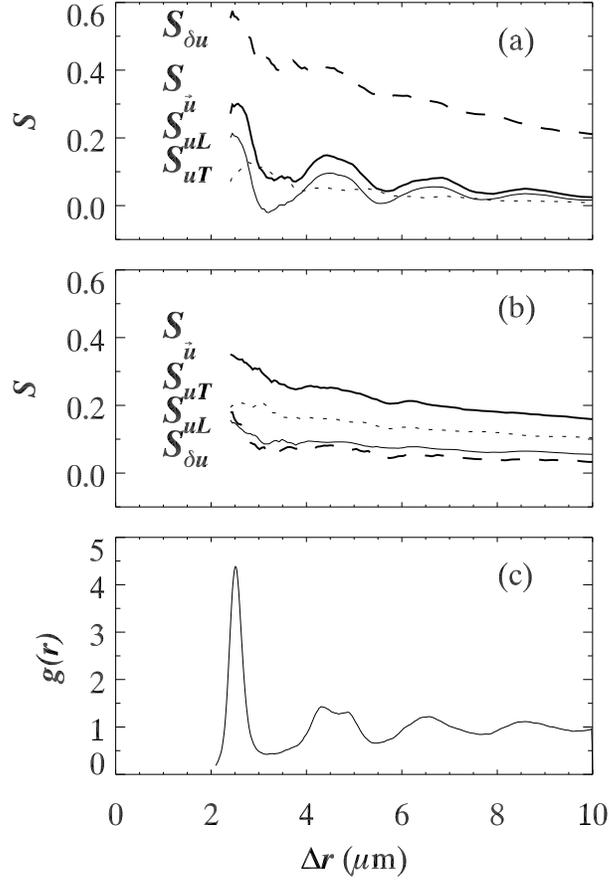}}
\smallskip
\caption{(a) Correlation functions for $\phi=0.56$ (a supercooled
fluid close to the glass transition), with $\Delta t = 2000$ s.
(b) Correlation functions for $\phi=0.62$ (a
glass), with $\Delta t = 10000$ s.
(c) $g(r)$ for the data shown in (a); $g(r)$ is similar for
the data shown in (b).}
\label{fig-dr}
\end{figure}

The results are harder to interpret for glassy samples (thin
lines in figure \ref{fig-dt}), due to the aging of the system.
In an aging system, the properties of the samples depend
on the time since the sample was prepared \cite{courtland}; for example, the
increase in $\langle \delta x^2 \rangle$ at large $\Delta
t$ [figure \ref{fig-dt}(a)] moves to longer $\Delta t$ as
the sample age increases.  (The sample age is defined as
the time since stirring the sample prior to the start of
the experiment.  To minimize the effects of aging for these
data, data acquisition was started 5-10 hours after stirring
the sample, which reliably initiates the aging \cite{courtland}.)
As seen in figure \ref{fig-dt}(c,d), any $\phi$
dependence for the correlated behaviour in glassy samples
is unclear.  The correlation functions are nonzero due to
correlated motion of particles within cages; however, the
increases seen in $S_{\vec u}$ and $S_{\delta u}$ at large
$\Delta t$ are probably due to the slight motions responsible
for aging \cite{courtland}.

To examine the spatial character of the highly correlated
particles occurring at the cage-rearrangement time scales,
we study the $\Delta r$ dependence of these functions fixing
$\Delta t$ to maximize $S_{\delta u}$ [figure \ref{fig-dt}(d)].
The correlation functions are shown for a supercooled fluid
($\phi=0.56$) in figure \ref{fig-dr}(a), indicated by the
thick lines.  Both functions oscillate, with especially
strong oscillations seen in $S_{\vec u}$ (thick solid line).
Strikingly, these oscillations coincide with oscillations of the
pair correlation function $g(r)$, shown in figure \ref{fig-dr}(c).
For example, pairs of particles with separations corresponding
to the first peak in $g(r)$ are nearest neighbors, and are
the most strongly correlated.  This agrees with our earlier
work which found that the pair correlation function $g(r)$
influences the motion of pairs of particles \cite{weeks01}.
In fact, comparing $S_{\delta u}$ and $S_{\vec u}$ shows that
the correlation of the {\it mobility} of pairs of particles
is less sensitive to $g(r)$, whereas the correlation of
their {\it directions} is more sensitive to $g(r)$.  This is
further evidence that cage rearrangements involve regions of
particles with high mobility, although within those regions
the directions of the displacements of individual particles
are not as strongly correlated, but rather are influenced by
$g(r)$.  Both simulations of Lennard-Jones particles and of
hard spheres saw similar oscillations in spatial correlation
functions, but the relationship with $g(r)$ was unclear
as the systems had higher polydispersity than our samples
\cite{donati99,doliwa00,cs}.

To further examine the correlations in the directions of motion
of pairs of particles, we decompose $S_{\vec u}$ into longitudinal correlations
(correlations of the displacement along the direction of the separation
vector $\Delta \vec r_{ij}$)
and transverse correlations (correlations of the
component of the displacement vector perpendicular to $\Delta \vec r_{ij}$).
Two new functions are defined as:
\begin{eqnarray}
S_{\vec u L}(\Delta r,\Delta t) &=& 
{\langle u_i^L u_j^L \rangle \over \langle u^2 \rangle}\\
S_{\vec u T}(\Delta r,\Delta t) &=& 
{\langle \vec u_i^T \cdot \vec u_j^T \rangle \over 
\langle u^2 \rangle}
\end{eqnarray}
where $u_i^L = \vec u_i \cdot \hat R_{ij}$,
$u_j^L = \vec u_j \cdot \hat R_{ij}$,
$u_i^T = \vec u_i - u_i^L \hat R_{ij}$,
and $u_j^T = \vec u_j - u_j^L \hat R_{ij}$.
The normalization is chosen so
that $S_{\vec u} = S_{\vec u L}+S_{\vec u T}$.  $\hat R_{ij}$ is
a unit vector pointing from particle $i$ to particle $j$, and the dot
products for the $u^L$ terms are taken so that two vectors pointed
in the same direction correlate positively.  These two
functions are also plotted in figure \ref{fig-dr}(a) (thin lines).
The oscillations of $S_{\vec u}$ are almost entirely due to the
contribution from $S_{\vec u L}$ (thin solid line), indicating
that the longitudinal motion is most sensitive to $g(r)$.
In fact, the longitudinal correlations of the motions of
nearest neighbor particle pairs are suggestive of ``string-like''
motion seen in simulations \cite{donati98}.  However, $S_{\vec u
L}$ also shows a slight anticorrelation at $\Delta r \approx
3.2$ $\mu$m, corresponding to the first minimum of $g(r)$.
This indicates that pairs of particles separated by $\approx
1.5\sigma$ are actually more likely to move in antiparallel
directions (towards or away from each other), despite their
strongly correlated mobility.
The transverse component $S_{\vec u T}$ (thin dotted line)
has a slight increase closer to the first minimum of $g(r)$,
but otherwise shows little dependence on $g(r)$.

\smallskip
\begin{figure}
\centerline{
\epsfxsize=8truecm
\epsffile{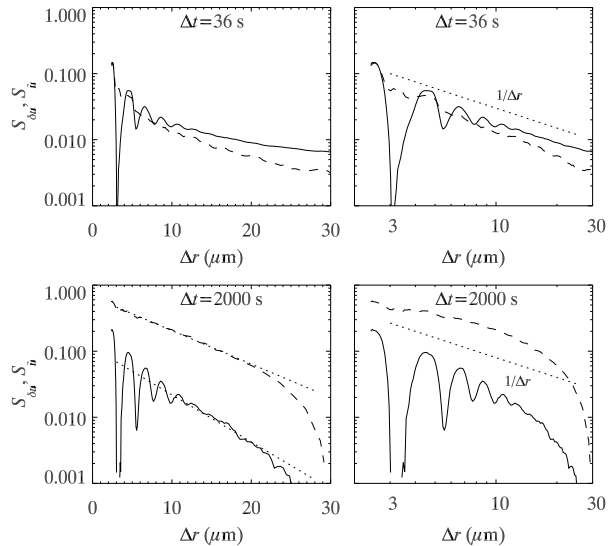}}
\smallskip
\caption{Semilog and log-log plots of $S_{\vec u}$ (solid lines) and 
$S_{\delta u}$ (dashed lines) at two different $\Delta t$
as indicated, for $\phi=0.56$, a supercooled fluid.
The dotted $1/\Delta r$ lines are drawn as a guide to the eye
in the log-log plots.  The dotted fit lines shown in the lower left
plot have decay lengths $\xi_{\delta u}=8.5$~$\mu$m and
$\xi_{\vec u}=6.1$~$\mu$m.  The downturn at long $\Delta t$ is
not fitted, and is described in the text.
}
\label{fig-bigdr}
\end{figure}

At larger separations, particle motion is still less correlated.
If the particles were in a dilute suspension and long range
interactions were solely due to the hydrodynamic behaviour
of the solvent, the correlation functions should decrease as
$1/\Delta r$.  This would also be the case for a homogeneous
viscoelastic medium \cite{crocker00}, although the localized
rearrangements observed previously shows that motion in these
dense colloidal samples is spatially inhomogeneous \cite{weeks00,kegel00}.
To check this, we plot the correlation functions on semilog
and log-log axes in figure \ref{fig-bigdr} for $\phi=0.56$ at
two different time scales.  The top right graph shows that the
decay is close to $1/\Delta r$ for short time scales; we find
similar behaviour for all liquid samples at time scales $\Delta
t < 100$~s.  The behaviour characteristic of longer time scales
relevant for cage rearrangements is shown in the bottom graphs.
On the semilog graph, the decays follow straight lines for
both functions out to $\Delta r \approx 25$~$\mu$m, indicating
exponential decay with characteristic length scales $\xi_{\delta
u} = 8.5$~$\mu$m = $3.6\sigma$ and $\xi_{\vec u} = 6.1$~$\mu$m
= $2.6\sigma$.  For this sample, the decay lengths vary in the
range $\xi_{\delta u} = 7.6 \pm 0.9$~$\mu$m and $\xi_{\vec u} =
6.7 \pm 0.9$~$\mu$m when $\Delta t$ is varied from 50~-~5000~s.
The exponential decay demonstrates that for dense colloidal
suspensions on these length and time scales are not behaving
as continuum viscoelastic materials \cite{levine00a,levine00b}.

At large $\Delta r$, a downturn is seen in the correlation
functions (bottom graphs in figure \ref{fig-bigdr}).  This is
due to a counterflow, similar to what was seen in simulations
\cite{doliwa00,alder70}.  This counterflow has been previously interpreted as
the medium's response to the transient motion of a particle,
although it is less clear how this may apply in the case of
localized motions \cite{doliwa00,alder70}.
The counterflow may be due to the
presence of the coverslip ($>$30~$\mu$m away), although the
correlation functions do not change significantly when the data
is split into two subsets, one closer to the coverslip and one
farther.  The counterflow results in a slight anticorrelation
in both $S_{\vec u}$ and $S_{\delta u}$ at $\Delta r \approx 40$
$\mu$m (not shown).  In all cases the exponential decay occurs
over a decade in $\Delta r$ before the counterflow cuts off
the correlation.  At still larger length scales ($\Delta r
\gg \xi_{\vec u}, \xi_{\delta u}$), the correlation functions must decay
as $1/\Delta r$, as on a sufficiently large length scale the
colloidal suspension will appear homogeneous.
However, we do not
have data at large enough $\Delta r$ to see $1/\Delta r$ decay,
and at large separations the amplitude of the correlation functions would be
quite small and difficult to measure.

\smallskip
\begin{figure}
\centerline{
\epsfxsize=8truecm
\epsffile{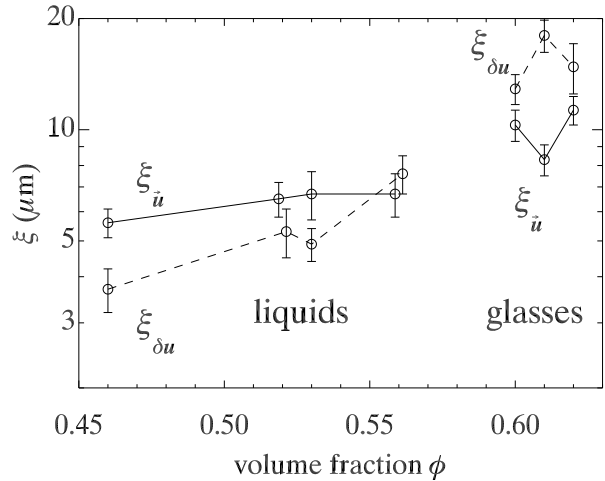}}
\smallskip
\caption{Volume fraction $\phi$ dependence of the dynamic length
scales, for colloidal liquids and glasses.  The length scales
are determined by a fit to the exponential decay of correlation
functions such as those shown in figure \protect\ref{fig-bigdr};
the error bars indicate variability in determining the length
scales at different $\Delta t$'s.  (The data points for
$\phi=0.52$ and $\phi=0.56$ are horizontally offset from
each other slightly for clarity.)
}
\label{fig-phi}
\end{figure}

To look for a possible growing length scale, we extract the
correlation decay lengths for $S_{\vec u}$ and $S_{\delta u}$
in samples with different volume fractions, with results shown
in figure \ref{fig-phi}.  The decay lengths increase only
slightly as $\phi_g$ is approached.  Within the error bars,
$\xi_{\vec u}$ is consistent with a constant value $\xi_{\vec
u} \approx 3\sigma$.  $\xi_{\delta u}$ doubles over the
range $\phi=0.46 - 0.56$, and this is sufficient to account
for the increase in the size of clusters of mobile particles
seen seen in previous work \cite{weeks00}; however,
we see no evidence in our data for
or against a divergence of $\xi_{\delta u}$ at $\phi_g$.
Intriguingly, this slight increase in $\xi_{\delta u}$ gives
new insight into the $\phi$ dependence of $S_{\delta u}$
for the supercooled fluids, seen in figure \ref{fig-dt}(d).
If we assume $S_{\delta u}(\Delta r) = A_{\phi}(\Delta t)
\exp(-\Delta r/\xi_{\delta u})$, we find that the maximum
value (with respect to $\Delta t$)
of $A_{\phi}(\Delta t)$ is a constant, approximately
$0.75 \pm 0.05$, independent of $\phi$.  Thus, the increasing
height of $S_{\delta u}$ as $\phi$ increases, seen in
figure \ref{fig-dt}(d), is primarily due to the increase
in $\xi_{\delta u}$ and thus an increasing value of
$\exp(-\Delta r_{\rm nn}/\xi_{\delta u})$, using the nearest
neighbor separation $\Delta r_{\rm nn}$.



We can compare our length scales $\xi_{\vec u}$ and $\xi_{\delta
u}$ with length scales obtained from computer simulations of
glass-forming systems.
The values we find for $\phi=0.56$ ($\xi_{\vec u} \approx
\xi_{\delta u} \approx 3\sigma$) are slightly larger than that
seen in simulations of hard spheres ($\xi_{\vec u} = 2.3\sigma,
\xi_{\delta u} = 1.4\sigma$); the simulations had a higher
polydispersity, and thus may have been farther from the glass
transition at $\phi=0.56$ \cite{doliwa00}.  Simulations of
Lennard-Jones particles did not notice exponential decay, and
instead computed a length scale by calculating (in our notation)
\begin{equation}
\xi'(\Delta t) = 
\left( {\langle u^2 \rangle \over \langle u \rangle^2} - 1\right)
\int_{0}^{\infty} d\Delta r \, S_{\delta u}(\Delta r,\Delta t) .
\end{equation}
The term in parenthesis is qualitatively similar to the
nongaussian parameter $\alpha_2$, and over the range of liquids
shown in figure \ref{fig-phi} varies from 0.35 (at $\phi=0.46$)
to 0.70 (at $\phi=0.56$).  We calculate $\xi'$ for our data by
using the form $A_{\phi}(\Delta t) \exp(-\Delta r/\xi_{\delta
u})$ for $S_{\delta u}$ with our measured values for $A$
and $\xi_{\delta u}$, and using the value of $\Delta t$
which maximizes $S_{\delta u}$ (and thus maximizes $\xi'$).
We find that $\xi'$ increases from 0.9 - 4.1 $\mu$m $\approx
0.4\sigma - 1.7\sigma$ as the glass transition is approached.
These values are larger than those seen in the simulation
(0.05 - 0.32 in Lennard-Jones units) \cite{poole98}.

Earlier work has seen evidence that structural properties
of the sample are slightly correlated with particle mobility
\cite{weeks01,conrad}.  This then suggests that the exponential
length scales seen in the correlation functions may relate
to spatial correlations of structural properties.  We have
computed a spatial correlation function $S_{\delta V}$ similar
to $S_{\delta u}$ (3), except looking at fluctuations
$\delta V$ of the Voronoi volume $V$ of each particle.
The Voronoi volume is a geometric way of partitioning space,
so that each particle claims the volume that is closer to its
center than to the center of any other particle.  Particles
with slightly larger Voronoi volumes are, in some sense,
seeing a slightly smaller local volume fraction, and this is
correlated with a slightly enhanced mobility \cite{weeks01}.
The correlations $S_{\delta V}$ do decay exponentially (not
shown), with a length scale of $5.7 - 6.3$ $\mu$m for the four
liquid samples we consider ($\sim 2.5 \sigma$).  Strikingly, we
see no clear volume fraction dependence of this length scale.
Furthermore, this length scale is quite similar to the length
scales seen for $S_{\vec u}$.  The lack of a clear connection
to the length scale for mobility ($\xi_{\delta u}$) suggests
that the correlation between the structure and the mobility,
while present, is not the entire story; the cooperative
motions are influenced by more factors than just the local
volume \cite{conrad,harrowell04}.

For the glassy samples, the behaviour is strikingly different.
As discussed before, our glassy samples are aging, although for
the duration of the experiment, little aging occurs.  (We
consider samples with a large age $t_w=3\cdot 10^4$~s
and time scales $\Delta t \le t_w$.)
For our data, and for lag times $\Delta t < 5000$~s, almost
all particle motion consists of particles moving back and
forth within their cages.  $S_{\vec u}$ and $S_{\delta u}$
have a low amplitude, as shown in figure \ref{fig-dt}(c,d)
and figure \ref{fig-dr}(b), indicating that most of the motion
is uncorrelated.  For the glassy samples, $S_{\vec u} >
S_{\delta u}$, similar to the liquids at small $\Delta t$.
Figure~\ref{fig-dr}(b) also shows that the correlation functions
oscillate much less, and thus the local structure [$g(r)$]
has a only a minor role for glasses.  Another intriguing
difference is that $S_{\vec u T} > S_{\vec u L}$, indicating
fluctuations in the transverse directions are more strongly
correlated than in the longitudinal direction.

The correlated motion that is present in the glasses, while
of small amplitude, is rather long-ranged.  Similar to the
supercooled fluids, the correlation functions for glasses show
exponential decay over a range of time scales $\Delta t$.  The
length scales for the decay are plotted in figure \ref{fig-phi}.
They are significantly larger than the length scales associated
with the supercooled fluids; for the glasses, $\xi_{v,\delta
v} = 8-20$ $\mu$m $\approx 3.5\sigma - 8.5\sigma$.  The data
shown in figure \ref{fig-phi} also indicate that the scalar
correlation lengths $\xi_{\delta u}$ are noticeably larger than
the vector correlation lengths $\xi_{\vec u}$, a trend which
is the opposite that of the liquids.  The existence of this
long-range correlation is sensible, given that the glasses
are more densely packed:  while most motion is localized
and uncorrelated, particles can move slightly if there is
long-range cooperation, even if they then move back (in order
that there be no rearrangements).  At present, however, it is
unclear if these correlation length scales correspond to the
intrinsic motion of the frozen glassy system, or if they are
connected with the aging process \cite{courtland}.  The previous
observations that clusters of ``mobile'' particles are smaller
is consistent with the low amplitude of the correlations
\cite{weeks00,courtland}.

Unlike the correlated motion in glasses, the structural
correlations in the glassy samples are not long-ranged.
The length scale for $S_{\delta V}$, the correlation function
of the fluctuations of Voronoi volumes, stays comparable to
that of the liquids ($\xi_{\delta V} \sim 5.9-7.7$ $\mu$m for
the glasses).  While there is very little change of this
length scale with $\phi$, we do note that the two samples
with the largest values of $\xi_{\delta V}$ correspond to the
same two samples with the largest values of $\xi_{\vec{u}}$,
again suggesting a connection between the structure and 
correlated particle motion \cite{cianci06}.

\section{Conclusions}

By measuring the positions of several thousand particles
over several hundred time steps, we have found that particle
motion is correlated over distances of 3-4 particle diameters.
In particular, particles undergoing cage rearrangements
have anomalously large displacements, and these are highly
spatially correlated.  We note that the largest degree of
correlation exists at the time scales corresponding to cage
rearrangements, but that our observations do not preclude
the sample from acting differently at very long time scales.  All
of our data is taken at time scales $\Delta t < \tau_\alpha$, so
we are not able to quantify the behavior on time scales
corresponding to alpha relaxation.

At time scales similar to the cage rearrangement time scale, the
correlations decay exponentially with separation $\Delta r$.
At very large $\Delta r$, we would
expect the correlations to recover a $1/\Delta r$ dependence as
predicted for homogeneous viscoelastic media
\cite{levine00a,levine00b}, at any
time scale.  This dependence is seen at short time scales in our
data (figure \ref{fig-bigdr}), suggesting that the instantaneous
response may be more continuum-like \cite{elastic}.  This is
sensible; in such short time scales, $\Delta t \ll \tau_\alpha$,
and so these motions are not related to dynamical
heterogeneities, and do not lead to flow of the sample.  Rather,
these are simple small amplitude affine deformations.  It is
intriguing that on the cage-rearrangement time scales, our
image volume is not large enough to see the $1/\Delta r$
behavior, suggesting that the appropriate coarse-graining length
scale is bigger than $\sim 25\sigma$.

The differing behaviour of the two correlation functions we
examine suggests that these rearrangements are composed
of regions of mobile particles (particles with large
displacements), but that the motions involve particles moving
in many directions, rather than all of the particles moving
in one coherent direction.  This is especially clear as the
correlations between the directions of motion of particles
($S_{\vec u}$) strongly depend on the pair correlation
function; within a group of mobile particles, their 
motions are often strongly directionally correlated, but also frequently
not directionally correlated.  This is consistent with
particles moving in necklace-like loops, for example, as seen
in Ref.~\cite{donati98}, and the mixing motions described in
Ref.~\cite{weeks01}.

The correlations become increasingly long-ranged as the
glass transition is approached, as seen in two new correlation
lengths $\xi_{\delta u}$ and $\xi_{\vec u}$.  This suggests that
rearrangements involving regions consisting of a small number of
particles become difficult or impossible as the volume fraction
increases, which would explain the growth of the viscosity as
the glass transition is approached.  The size of these regions,
quantified by the two correlation lengths $\xi_{\vec u}$ and
$\xi_{\delta u}$, are direct experimental evidence
for dynamical length scales near the glass transition.

\ack

We thank L.~Berthier, B.~Doliwa, U.~Gasser, and S.~C.~Glotzer
for helpful discussions.  We thank A.~Schofield for providing
our colloidal samples.  This work was supported by NSF
(DMR-0239109, DMR-0602584) and the Harvard MRSEC (NSF
DMR-0213805).

\vspace{0.5cm}




\begin{thebibliography}{99}
\bibitem{cv}  Our function $S_{\vec u}$ can be related to the
two-particle function described in \protect\cite{crocker00}:
\begin{equation}
D_{\alpha,\beta}(\delta r,\Delta t) = \langle u_i^{\alpha} u_j^{\beta}
\rangle
\end{equation}
where $\alpha$ and $\beta$ indicate the components of the vector
displacements to multiply.  Thus $D(\delta r,\Delta t)$ can be related to
$S_{\vec u}(\Delta r,\Delta t)$ as
\begin{equation}
D_{xx}+D_{yy}+D_{zz} =
S_{\vec u}(\Delta r,\Delta t) \langle |u|^2 \rangle .
\end{equation}
Note that $\langle |u(\Delta t)|^2 \rangle$ is just the ordinary
mean square displacement, and is $\Delta r$ independent,
so to convert from $S_{\vec u}(\Delta r,\Delta t)$ to $D(\Delta
r,\Delta t)$ is simple.
\bibitem{cs} Our function $S_{\delta u}$ can be related to the
two-particle function described in \protect\cite{donati99},
$g_u(\Delta r,\Delta t)$, which is
analogous to the static pair correlation function $g(r)$.
The static pair correlation function is defined by
\begin{equation}
g(\Delta r) = {1 \over 4 \pi (\Delta r)^2 \langle n \rangle N}
\sum_{i,j \neq i} \delta(\Delta r - R_{ij})
\end{equation}
which is normalized so that
$g(\Delta r) \rightarrow 1$ as
$\delta r \rightarrow \infty$.  $N$ is the number of particles in the sample,
$\langle n \rangle$ is the number density, and $\delta()$
is the Dirac delta function.  Using our notation,
\begin{eqnarray}
g_u(\Delta r,\Delta t) &=&
{{\left\langle \sum_{i,j \neq i}
u_i u_j \delta (\Delta r-R_{ij}(t)) \right\rangle_t }
\over
{4 \pi (\Delta r)^2 \langle u \rangle^2 \langle n \rangle N}}
\\
&=&
{\langle u_i u_j \rangle \over \langle u \rangle^2} g(\Delta r)
\end{eqnarray}
where the angle brackets $\langle \rangle_t$ around the sums
in the first line indicate a time average, and $\langle u_i u_j \rangle$
is taken over all pairs of particles with separations $\Delta r$.  
It can be shown that
\begin{equation}
{g_u(\Delta r,\Delta t) \over g(\Delta r)} - 1 =
 \left( {\langle u^2 \rangle \over
\langle u \rangle^2} - 1\right) S_{\delta u}(\Delta r,\Delta t)
\end{equation}
thus relating our $S_{\delta u}$ with the function $g_u$; the left hand
side of this equation is equivalent to $\Gamma(\Delta r,\Delta t)$
discussed in \protect\cite{donati99}.
$g_u$ is proportional to $g(\Delta r)$, whereas $S_{\delta u}$ is not.
\end{thebibliography}
\end{document}